\documentclass[15pt]{article}
\usepackage{indentfirst}
\usepackage{amsmath}
\usepackage{graphicx}
\usepackage{hyperref}
\usepackage{amssymb}

\author{Stanislav Srednyak}

\title{Properties of solutions of the "naive" functional Schroedinger equation for QCD.}
\begin{document}
\maketitle

\abstract{In this paper we consider the simplest functional Schroedinger equation of a quantum field theory (in particular QCD) and study its solutions. We observe that the solutions to this equation must possess a number of properties. Its Taylor coefficients are multivalued functions with rational and logarithmic branchings and essential singularities of exponential type. These singularities occur along a locus defined by polynomial equations. The conditions we find define a class of functions that generalizes to multiple dimensions meromorphic functions with finite Nevanlinna type. We note that in perturbation theory these functions have local asymptotics that is given by multidimensional confluent hypergeometric functions in the sense of Gelfand-Kapranov-Zelevinsky.

\section{Introduction}


In this paper we derive some basic properties of solutions of the simplest version of functional Schroedinger equation in quantum field theories. There is little doubt that a version of functional Schroedinger equation (FSE) underlies dynamics of bound states in QCD. However, there are several possible such equations. One popular version is being actively investigated in the framework of light cone quantlization ~\cite{Brodsky}. Other versions are possible. In this paper we concentrate on the most naive such equation, the one that comes from second quantization in canonical formalism. In the simplest case of 3-dimensional $\phi^3$ theory it can be written as 
\begin{multline}
\label{eq:1}
\partial_t F(t)[a_p]=H(t)[a,\delta/\delta a_p] F=\\
=(\int K_{3\leftarrow 0}(p,q,r)e^{it(-\omega_p-\omega_q-\omega_r)}\delta^3(p+q+r)a_pa_qa_r+\\
K_{2\leftarrow 1}(p,q,r)e^{it(-\omega_p-\omega_q+\omega_r)}\delta(p+q-r)a_pa_q\delta_r+\\
K_{1\leftarrow 2}(p,q,r)e^{it(-\omega_p+\omega_q+\omega_r)}\delta(p-q-r)a_p\delta^2_{qr}+\\
K_{0\leftarrow 3}(p,q,r)e^{it(\omega_p+\omega_q+\omega_r)}\delta(-p-q-r)\delta^3_{pqr})F(t)[a_p]
\end{multline}
This equation is formulated on the "manifold" of quantum fields $a_p=a(p),p \in C^3$, which is the function space of continuous functions of 3-momenta $p$, and $\delta_p=\delta/\delta a_p$ is the functional derivative (annihilation operator of second quantization). Note that the Hamiltonian $H(t)[a_p,\delta_p]$ is time dependent. This is essential feature of QFTs that this dependence is nontrivial. It reflects the fact that bound states in QFT have non-trivial internal dynamics, i.e., there are particle creation/annihilation processes in the virtual cloud surrounding the bound state. We give a more detailed discussion of this equation in section 2 below, and exhibit the Hamiltonian for QCD in the appendix.

We want to emphasise at this point that the above Schroedinger formulation is probably not the right one. It is quite difficult to incorporate the complicated geometry of particle collisions ~\cite{tree} into this framework. Probably, the right equation would be of the sort 
\begin{equation}
\delta/\delta t(x) F(x)[a_p]=H(\xi(x))[a_p,\delta/\delta a_p]F(x)[a_p]
\end{equation}
where the time $t(x)$ is local. The solution than would allow to "quantize" the geometry of multiparticle events ( by quantization I mean here determination of probabilities of particular channels of the reaction in an exclusive process, and restriction on the number and momentum space geometry of such channels). Such equations were at the beginning of quantum field theory ~\cite{Tomonaga}, but now are no longer studied because of the complexity of their solutions. We believe however that it is very important to continue the study of such equations, both for physical and for mathematical reasons. 

The purpose of this article is more modest, however. We will concentrate on the above "naive" version of the Schroedinger equation, with one single time, and study properties of its solutions. These properties follow almost immediately from the form of the Hamiltonian. They point out to a class of holomorphic functions ( in fact, sections of a vector bundle) defined on the space of complexified momenta $\{(p_1,...,p_n)\}=\mathbb{C}^{3n}$. We think that this is a very interesting class of functions. It is a multidimensional generalization of what is known as meromorphic functions of finite Nevanlinna order ~\cite{Nevanlinna_PDE, Nevanlinna_error_term,Nevanlinna_Vojta}. In particular, they generalize the exponential function, in a non-trivial way. We show that , locally, these functions are multidimensional confluent hypergeometric functions, in the sense of Gelfand-Kapranov-Zelevinsky ~\cite{GZK_toric,GZK_Dmod}. Apart from essential singularities, these functions possess regular singularities, the ones that locally are captured by holonomic D-modules with regular singularities. One technical difficulty, that arises in summation of infinitely many terms in perturbation theory, is that we will need to allow explicitely infinite dimensional bundles. We overcome this difficulty by constructing a filtration on the local monodromy and on the essential (Stokes) , asymptotic data that determine the functions $F_\alpha(p_1,...,p_n)$. We observe that this natural class of solutions is significantly more complicated than the one studied in ~\cite{Brodsky}, which was one of the motivations for this paper. 

We note that the Schroedinger equation generalizes to open valence states. It is natural to consider wave functionals that have explicit dependence on a tuple of points in the momentum space. For example, for 3-valent state (baryon) we can consider the wave functional $F(t,p,q,r)[a_s]$. The Schroedinger equation has similar form
\begin{equation}
\partial_t F(t,p,q,r)[a_s]=H(t)[a_s,\delta_s]F(t,p,q,r)[a_s]
\end{equation}
It is natural to ask if there are anlogies of Dirac equation in the functional space, i.e., if it is possible to consider nontrivial functional bundles, and allow for explicit dependence of the differential operator on the momentum space point. Indeed, such generalizations exist. We hope to consider them in future publications. 

We give basic formulation of the problem in Sec. 2. Sec. 3 contains our main results. In the Appendix we present the generalization to QCD.

\section{The "naive" functional Schroedinger equation for scalar QFT}


In this section we discuss the motivation for considering the functional Schroedinger equation ~\ref{eq:1} and its versions. This equation follows from the most basic formulation of second quantization in QFT ~\cite{Dirac_QFT,Bogoliubov_QFT}. The creation and annihilation operators $a_p,a^+_p$ were introduced by physicists in 1930's ~\cite{Dirac_QFT} in the attempt to model systems with the changing number of particles. What is not widely appreciated is that these operators can be interpreted as multiplication by a function and variational derivative with respect to a function variable. This leads immediately to a version of differential topology on functional manifolds. The basic starting point of the scattering theory is the following expression for the S-matix 
\begin{equation} 
S=Texp\int_{-\infty}^{+\infty} H_{int}(t)dt
\end{equation}
This S-matix can be used to obtain perturbative expressions for e.g. triangle diagram, which then leads to the well known expressions of anomalous magnetic moment. It is remarkable fact that the expressions derived for anomalous magnetic moment of the electron agree with experiment to high order in perturbation theory ~\cite{Kinoshita}. Feynman diagrams capture all the vacuum polarization that accounts for anomalous magnetic moment. The situation with bound states is far less clear. For example, already in the case of Lamb shift beyond one loop, it is impossible to formulate closed system of equations that would allow to calculate the spectrum beyod one loop ~\cite{Eides}. The problems that arise in bound state formulations are already visible in the "derivation" of Dirac equation with radiative correstions in ~\cite{Bogoliubov_QFT,Itzykson} or in the necessity to choose equal time formalism in spectral calculations ~\cite{Pachucki,NRQED}. An interesting modern approach is ~\cite{Shabaev}.

The purpose of this article is to introduce the class of solutions for the Schroedinger tower of multiparticle wave funtions. We will use the most classic Hamiltonian for illustration purposes. This Hamiltonian can be written in the form
\begin{equation}
H_{\phi^3}(t)[a_p,\delta_p]=\int d^3x \lambda (\int \frac{d^3p}{(2\pi)^{3/2}} (\frac{a_p}{2\sqrt{\omega_p}}e^{-i\omega_pt+ipx}+\frac{\delta_p}{2\sqrt{\omega_p}}e^{i\omega_pt-ipx})^3
\end{equation}
After some algebra, we arrive at eqn. ~\ref{eq:1}. We can be more explicit about the coefficients $K_{i\leftarrow j}(p_1,...,p_i;q_1,...,q_j)$ that can be considered as (parts of ) transition kernels of $j$ to $i$ elementary quanta. Note that these functions play a prominent role in the derivation of splitting functions for parton evolution equations of QCD ~\cite{Indian_boy}. In our simple case $i+j=3$, but our results concerning analytic structure of the solutions can be generalized to arbitrary $i+j$, in particular, to various effective thories. In the case of scalar field theory these functions equal
\begin{equation}
K_{i\leftarrow j}(p,q,r)=\frac{1}{\sqrt{\omega_p \omega_q \omega_r}}
\end{equation} 
In the case of QCD, they also involve numerators, which are the wave functions of elementary quanta (see Appendix). The functions $K_{i\leftarrow j}(p,q,r)$ are algebraic functions of their arguments. They are holomorphic branching functions, once analytic continuation of momenta into the complex space was performed. In the following, our momenta will take values in the complex 3-dimensional space $\mathbb{C}^3$. The functions $K_{i\leftarrow j}(p,q,r)$ are finite branched covers of $\mathbb{C}^9$ that branch on an algebraic variety (in fact, a set of conics). 

The wave functional $F[a_p]$ contains the information about the multiparticle wave functions
\begin{equation}
\label{eq:F}
F[a_p]=\sum_{n=0}^{\infty} \int dp_1...,dp_n F(p_1,...,p_n)|p_1,...,p_n>
\end{equation}
It belongs to the Fock space of the theory ~\cite{Fock}. Note that Fock space is the subject of much research in the theory of $C^*$ and von Neumann algebras ~\cite{Shlyakhtenko}( and references therein), in particular, of $C^*G$, for discrete groups $G$ ~\cite{Phillips} ( in particular, for $G=\pi_1M$, fundamental groups of a manifold $M$). The Fock space of a QFT ( the container of all bound states, in particular, heavy nuclei) has much more geometric structure that the Fock space in operator algebra theory (in particular, it is poly-normed in QFT case). However, we believe that there are many analogies between these two developments, and many fruitful parallels can be drawn (many of these are alredy studied in the vertex algebra litearature ~\cite{Borcherds,Jones}).

\section{Basic facts about solutions of functional Schroedinger equation}

In this section we collect some very basic observations about perturbative solutions of the functional Schroedinger equation. In particular, we state conditions for existence of periodic solutions. We start with preliminary observation that limits the simplicity of functional forms of the solutions to bound state problem.

{\bf Observation 0.} Solutions to FSE that correspond to stable bound states have nontrivial (nonfactorizable) time dependence. There is internal dynamics inside QFT bound states. 

This observation is generic and is valid in all models of bound states. Our naive model has just one time variable. However the conclusion must hold in more advanced models with multidimensional time. 

To motivate our discussion, we start with perturbation theory for the FSE. Suppose , for simplicity, that at time zero the functional had expansion ~\ref{eq:F} such that all the components $F(p_1,...,p_n)$ are rational functions of momenta. The conclusions will be valid for more general choice of $F(t=0)$ that can contain exponetals of algebraic (in broad sense) functions, and also "generalized exponential functions"( see definition below). This class of functions contains all examples that author ever saw in physics literature. Consider then the following n-th term in time ordered perturbation theory 
\begin{equation}
S_n(t,0)=\int_0^t dt_1 H(t_1) \int_0^{t_1}dt_2 H(t_2)...\int_0^{t_{n-1}}dt_n H(t_n) F
\end{equation}
The expression for $H(t_1)...H(t_n)F$ can be obtained using a version of Wick theorem. The result is similar to a sum of Feynman diagrams, with modified, time-dependent vertices. The evolution operator can be written symbolically as 
\begin{multline}
S(t,0)=\sum_{a,b,c} A_{a,b,c}(p_1,...,p_a,q_1,...,q_b,s_1,...,s_c)\times \\
\times e^{t\omega_{a,b,c}(p_1,...,p_a,q_1,...s_1,...)} a_{p_1}...a_{p_a}\delta^b_{q_1,...,q_b} +\\
\sum_{a,b,c} B_{a,b,c}(p_1,...,p_a,q_1,...,q_b,s_1,...,s_c) a_{p_1}...a_{p_a}\delta^b_{q_1,...,q_b}
\end{multline}
In this expression, $A_{a,b,c},B_{a,b,c},\omega_{a,b,c}$ are certain algebraic functions of the arguments $(p_1,...,p_a,q_1,...,q_b,s_1,...,s_c)$. These functions can be expressed through roots of polynomials. They are finite branched covers of the affine space $\mathbb{C}^{3(a+b+c)}$ that branch on divisor containing finite set of algebraic varieties. Functions $\omega_{a,b,c}(p_1,...,p_a,q_1,...,q_b,s_1,...,s_c)$ have quite substantial structure. They can be expressed through linear combinations of elementary frequency functions taken at a linear combination of 3-momenta $p_i,q_j,s_k$. The functions $A,B$ are products of the splitting functions $K_{i\leftarrow j}$ and the usual factors $1/(\pm\sqrt{\omega_l}\pm \sqrt{\omega_n}\pm \sqrt{\omega_m}+i\epsilon)$. 

Suppose now that we apply the evolution operator $S(t,0)$ to a generiac rational functional $F(t=0)$. Then we claim that the functional $F(t)$ will have the following properties

{\bf Property 1.} $F_{p_1,...,p_n}$ has regular singularities along a divisor $D=\cup_{i=0}^{\infty} D_i$ in the space $(p_1,...,p_n)$ that has infinitely many components $D_i$ whose degree can grow arbitrarily large.

{\bf Property 1'.}  $F_{p_1,...,p_n}$ has essential singularities along a divisor $D'=\cup_{i=0}^{\infty} D'_i$ in the space $(p_1,...,p_n)$ that has infinitely many components $D'_i$ whose degree can grow arbitrarily large. Each component of this divisor is an algebraic variety.

{\bf Property 2.} $F_{p_1,...,p_n}$ is a pro-finite bundle on the complement $D\cup D'$. In other words, $F_{p_1,...,p_n}$ has an index $\alpha$, above every point we have a vector space of values of $F_{p_1,...,p_n}$, there is a non-trivial monodromy around each $D_i$ and a non-trivial Stokes factor for each of the Stokes sectors associated with $D'_i$. The index $\alpha$ is not bounded, but both monodromy matrix and Stokes matrices are block-diagonal.

{\bf Property 3.} Near each essential singularity, $F_{p_1,...,p_n}$ is a "generalized exponential function".

We have to define what we mean by "generalized exponential function". 

{\bf Def. 1} By simple generalized exponential function we mean the following class of functions 
\begin{equation}
w(f,g)=\int f(p,q) e^{g(p,q)} d^nq
\end{equation}
Here, $p=(p_1,...,p_m),q=(q_1,...,q_n)$ are complex variables, and $f,g$ are algebraic functions of $p,q$. For example they can be rational or involve roots. But we also wish to include solutions to arbitrary algebraic equations, which leads to the class of algebraic hypergeometric functions ~\cite{Sturmfels_alg,Sturmfels_book}. These functions form a class of generalized (in the sense of Gelfand-Kapranon-Zelevinsky) confluent hypergeometric functions. In particular, all of them form the solution complex of certian algebraic holonomic D-module ~\cite{Reichelt, Walter}

{\bf Def. 2} By iterated generalized exponential functions we mean the class of functions 
\begin{multline}
w(f_T,...,f_1;g_T,...,g_1)=\\
=\int dq_T f_T(p,q_T)e^{g_T(p,q_T)}\int dq_{T-1} f_{T-1}(p,q_T,q_{T-1})e^{g_{T-1}(p,q_T,q_{T-1})}...\times\\
\times\int dq_1 f_1(p,q_T,...,q_1)e^{g_1(p,q_T,...,q_1)}
\end{multline}

The last definition attempts to capture the notion of "filtration" on the bundle $F_\alpha$ that is possessed by the perturbative expansion. Both monodromy and Stokes matrices are explicitely infinite after summation of all orders in perturbation theory, but they can be constructed from finite blocks. 

The proof that application of the perturbative time dependent evolution operator leads to the above class of functions is not difficult. In fact, our definitions were tailored to capture in as simple as possible terms the class of functions emerging in perturbative QFT. There are two facts that need to be proved. First, that there are infinitely many components in the singularity loci. This can be done by examining the usual pinch condition for the integrand. The general mechanism of emergence of singularities of holomorphic functions is the same for both rational integrands and the integrands containing the exponentials - there must be a topological transition in the branch ( or pole) locus. This leads immediately to discriminantal varieties ~\cite{GZK_book}. One only needs to observe that the intergand is sufficiently generic and that there are indeed infinitely many components in the singularity locus. Second, we need to demonstrate that the singular behaviour near essential singularity is governed by Stokes data. This follows from general theory of confluent hypergeometric functions ~\cite{GZK_Euler, Sabbah, Kedlaya, Boalch}. There is singinificant amount of structure in the functions arising in QFT (i.e. they are not generic), and they can be made more explicit.

\section{Conclusion}

We considered a version of functional Schroedinger equation that arises naturally in QFTs. We discussed some very basic properties of its solutions. Our set up was the most basic one, and we believe that all realistic solutions to the functional Schroedinger equation ( whichever version of it one chooses) must possess these properties. The properties point out to a rather extensive class of holomorphic functions of several variables, that in particular contains multidimensional entire functions of finite order of growth near essential singularities, and branching functions that generalize the classical polylogarithms and elliptic polylogarithms. This class of functions deserves further study.

\section{Appendix. Functional Schroedinger equation for QCD.}

For convenience of the reader, we present here the formula for the second-quantized Hamiltonian of QCD. For simplicity, we omit ghosts ( which can be reinstated following e.g. ~\cite{Popov}). The quantized gluon field comes in $N_c$-plet $g_{a,s,p}$, $a=1,...,N_c$, $s=\pm 1$ is the spin variable. The pure gauge part of the Hamiltonian is 
\begin{multline}
H_{YM}(t)[g_{a,s,p},\delta_{a,s,p}]=\\
=\sum_{k=3,4} \sum_{i,j=0..k,i+j=k} \sum_{a_r,s_r} K^{i\leftarrow j}_{(a_1,s_1,...,a_k,s_k)}(p_1,...,p_i;q_1,...q_j)\times\\
\times\delta^3(p_1+...+p_i-q_1-...-q_j)\times \\
e^{it(\omega_{p_1}+...+\omega_{p_i}-\omega_{q_1}-...-\omega_{q_j})}g_{a_1,s_1,p_1}...g_{a_i,s_i,p_i}\frac{\delta}{\delta g_{a_{i+1},s_{i+1},q_1}}...\frac{\delta}{\delta g_{a_{k},s_{k},q_j}}
\end{multline}
Here $K^{i\leftarrow j}$ are the usual $3,4$-gluon interaction vertices. For example, the 4-gluon vertex $K^{2\leftarrow 2}_{(a_1,s_1),...,(a_4,s_4)}(p,q,r,t)$ involves several terms of the form
\begin{multline}
K^{2\leftarrow 2}_{(a_1,s_1),...,(a_4,s_4)}(p,q,r,t)=\\
=-g^2\sum_{w,v,u,z}T^d_{a_u,a_v}T^d_{a_w,a_z}\frac{\epsilon_{s_w,\mu}(k_w)\epsilon_{s_v,\mu}(k_v) \epsilon^*_{s_u,\nu}(k_u)\epsilon^*_{s_z,\nu}(k_z)}{ \sqrt{\omega_{k_w}\omega_{k_v}\omega_{k_u}\omega_{k_z}}}
\end{multline}
where the sum is extended over all permutations $w,v,u,z$ of $p,q,r,t$ and $\epsilon_{s,p}$ is the gluon polarization tensor. It is important to note that this tensor is an algbraic function of $p$. The 3-gluon interaction has the form 
\begin{multline}
K^{1\leftarrow 2}_{(a_1,s_1),...,(a_3,s_3)}(p,q,r)=ig\sum_{w,v,u}\frac{\epsilon_{s_w,\mu}(k_w)\epsilon_{s_v,\mu}(k_v) k_\nu \epsilon_{s_u,\nu}(k_u)}{ \sqrt{\omega_{p}\omega_{q}\omega_{r}}}
\end{multline}
where $k=k_w$ or $k=k_v$ ( it cannot equal $k_u$). Fermion terms are obtained analogously. The fermion/antifermion creation operators correspond to the anticommuting fields $b_{i,s,p},c_{i,s,p}$ where $s=\pm 1/2$ and $i$ is the color index. It will be convenient to unite the fields $b,c$ into single field $d_{e,i,s,p}$, where $e=\pm$ denotes particle or anti-particle. Then the interaction Hamiltonian contains $8N_cD$ terms, linear in $g$ and $\delta/\delta g$ derivatives, and bilinear in $b_p,c_p$ and variational derivetives in these fields. One such terms is written below
\begin{multline}
H_{YM-fermion}^{f\leftarrow gf}(t)=ig\int K^{f\leftarrow gf}_{i,s_1,j_s2,a_s2}(p,q,r)\\
\delta^3(p-q-r)e^{it(\omega_p-\omega_q-\omega_r)} b_{i,s_1,p}\frac{\delta}{\delta g_{a,s_2,q}}\frac{\delta}{\delta b_{j,s_3,r}}
\end{multline}
where

\begin{equation}
K^{f\leftarrow gf}_{i,s_1,j,s_2,a,s_2}(p,q,r)=\frac{T^a_{ij}\bar{v}_{s_1}(p)\hat{\epsilon}_{s_2}(q)v_{s_3}(r)}{\sqrt{\omega_p\omega_q\omega_r}}
\end{equation}

\bibliographystyle{amsplain}
\bibliography{bibliogr}

\end{document}